\documentclass[aip,jcp,reprint]{revtex4-1}
\usepackage{physics}
\usepackage{enumerate}
\usepackage{chemformula}
\usepackage{subcaption}
\usepackage{siunitx}
\usepackage{bigstrut}

\renewcommand{\vec}{\mathbf}
\newcommand{\rr}{|\vec{r}-\vec{r}'|}
\newcommand{\rhoscr}{\rho_{\textrm{scr}}}

\draft % marks overfull lines with a black rule on the right

\begin{document}
%\showthe\columnwidth

%\preprint{}

%\title{A density inversion method to obtain Kohn-Sham potentials} %Title of paper
\title{Density-inversion method for the Kohn-Sham potential: role of the screening density}



\author{Timothy J. Callow}
\email{timothy.callow@durham.ac.uk}
%\thanks{}
\affiliation{Department of Physics, Durham University, South Road, Durham, DH1 3LE, United Kingdom}
\affiliation{Max-Planck-Institut f\"{u}r Mikrostrukturphysik, Weinberg 2, D-06120 Halle, Germany}

\author{Nektarios N. Lathiotakis}
\email{lathiot@eie.gr}
%\thanks{}
\affiliation{Theoretical and Physical Chemistry Institute, National Hellenic Research Foundation, Vass. Constantinou 48, 116 35 Athens, Greece}

\author{Nikitas I. Gidopoulos}
\email{nikitas.gidopoulos@durham.ac.uk}
%\thanks{}
\affiliation{Department of Physics, Durham University, South Road, Durham, DH1 3LE, United Kingdom}


\date{\today}

\begin{abstract}
We present a method to invert a given density and find the Kohn-Sham (KS) potential in Density Functional Theory (DFT) which shares that density. Our method employs the concept of screening density, which is naturally constrained by the inversion procedure and thus ensures the density being inverted leads to a smooth KS potential with correct asymptotic behaviour. We demonstrate the applicability of our method by inverting both local (LDA) and non-local (Hartree-Fock and Coupled Cluster) densities; we also show how the method can be used to mitigate the effects of self-interactions in common DFT potentials with appropriate constraints on the screening density.
\end{abstract}

\pacs{}% insert suggested PACS numbers in braces on next line

\maketitle %\maketitle must follow title, authors, abstract and \pacs

\section{Introduction}
%\label{}

Density functional theory (DFT) is the most widely-used method in electronic structure theory calculations, with many tens of thousands of publications using it every 
year \cite{pribram2015dft}. Despite the many successes of the Kohn-Sham (KS) formalism in DFT, the most commonly used functionals do not correctly describe 
various physical situations, such as molecular dissociation and charge transfer processes \cite{Wang_insights,Maitra_2017}. Developing methods to overcome these 
difficulties is an active area of research \cite{SCAN_Perdew,Gorling_RPA,Pitts2018,Barlett_adventures,Yang_deloc,Burke_HF_DFT}.

In order to judge the quality of new approaches in KS theory, it is important to have an accurate reference against which to benchmark results. Often, we can compare with experiment or a higher level calculation; however, it is also valuable to know what an `exact' KS result is. This is commonly done by inverting an accurate density to find the corresponding KS potential. {\color{black}Various methods have been developed to accurately obtain the KS potential from a given density. Early attempts typically focussed on small atomic systems \cite{Almbladh_pedroza, Aryasetiawan, Nagy_March_1, Nagy_March_2, Nagy_1, Nagy_2, Chen_Stott_inversion, Werden_davidson}; more generally applicable methods\cite{Wang_Parr,Parr_inversion, Gorling_inversion, WY_inversion, Leeuwen_inverse, Kumar_2019, kanungo2019}, including to the time-dependent case\cite{Hodgson_inverse,Nielsen_tdinversion,Wasserman_tddft}, have subsequently been developed. However, the problem remains interesting due to its associated difficulties \cite{Wasserman_review}}.

In this paper, we present a method \cite{Hollins_LFX} to invert a known target density $\rho_t$ of 
a system of {\color{black}$N$} interacting electrons in a {\color{black}known} external potential $v_\textrm{en}${\color{black}, in order to obtain the 
Hartree-exchange and correlation (Hxc) potential of the KS system with density $\rho_t$.} 
Our method is based on minimizing the Coulomb energy $U[ \rho_v - \rho_t]$ of the density difference 
$\rho_v - \rho_t$, 
% between the density $\rho_v $ %of the KS potential $v_{en}+v$ and $\rho_t $, 
\begin{equation} \label{eq:obju}
U [ \rho_v - \rho_t] = \frac{1}{2}\iint\dd{\vec{r}} \dd{\vec{r}'} \frac{[\rho_v(\vec{r}) - \rho_{t}(\vec{r})][\rho_v(\vec{r}') - \rho_{t}(\vec{r}')]}{|\vec{r}-\vec{r}'|}  ,
\end{equation}
where $\rho_v$ is the density of 
{\color{black} 
another noninteracting $N$-electron system with KS potential $v_{\rm en} + v$. %This noninteracting system is also a KS system of course. %e latter is also a KS system with density $\rho_v$. 
%We denote it by $v_s [\rho_v] = v_{\rm en} + v$. 
Obviously, the effective potential $v$ simulates the electronic repulsion and at 
the minimum of the Coulomb energy $U$, when $\rho_v = \rho_t$, this effective potential becomes equal to the 
Hxc potential we seek.\footnote{
%As long as $\rho_v \ne \rho_t$, the potential $v$ differs from the Hxc potential of the KS system with density $\rho_v $. 
A subtle point is that in general, $v$ is not exactly equal to the Hxc potential of the KS system with density $\rho_v$.
Since $v_{\rm en}$ is the external potential for the KS system with density $\rho_t$, it cannot also be 
the external potential for the KS system with density $\rho_v$. 
Hence, as long as $\rho_v \ne \rho_t$, the potential $v$ is not exactly equal to the Hxc potential of the KS system with density $\rho_v$.} }
%
%\footnote{Since $v_{\rm en}$ is the external potential for the KS system with density $\rho_t$, it cannot also be 
%the external potential for the KS system with density $\rho_v$, when $\rho_v \ne \rho_t$. 
%Hence, neither can $v$ be the Hxc potential of the latter KS system.}.

The Coulomb energy $U$ is clearly positive and tends to zero as the two densities become close. As will be explained in section \ref{Algorithm}, 
minimizing $U$ also minimizes the energy difference from Ref.~\onlinecite{Gidopoulos_1},
\begin{equation} \label{eq:Tdiff}
  T_{\Psi}[v] = \mel{\Psi}{H_v}{\Psi} - E_v,
\end{equation}
where $\Psi$ is a state with density $\rho_t$, {\color{black}and $H_v$ is the many-body KS Hamiltonian,
\begin{equation} \label{ksv}
    H_v = \sum_{i=1}^N \left[-\frac{\nabla_i^2}{2} + v_\textrm{en}(\vec{r}_i) + v(\vec{r}_i)\right] ,
\end{equation}
of the KS system with density $\rho_v$.
When $\Psi$ is the (exact or approximate) ground state of the interacting system in the {\color{black} external} potential  
$v_\textrm{en}$, the minimizing potential of (\ref{eq:obju}, \ref{eq:Tdiff}) will be equal to (exactly or approximately) the Hxc potential of the KS system with density $\rho_t$.}

Central to our method is the concept of screening density \cite{Gidopoulos_CLDA}, or electron repulsion density \cite{Pitts2018}, in the KS scheme. 
It can be thought of as the effective electron density that screens the nuclear charge from a KS electron (i.e. electron in a KS orbital). 
Alternatively, it is the effective charge density that repels each KS electron, mimicking the electron-electron repulsion and 
underpinning the Hartree, exchange and correlation (Hxc) potential. Specifically, using Poisson's equation, the screening density can be obtained from the
Laplacian of the Hxc potential, $\rho_{\rm scr} (\vec{r}) = - (1/4 \pi) \, \nabla^2 v_\textrm{Hxc} ({\vec{r}})$ \cite{Gidopoulos_CLDA,Pitts2018}. {\color{black}G\"orling \cite{Gorling_uncontracted} and Liu, Ayers and Parr\cite{Liu_parr} had previously considered the xc-only screening density,  obtained from the Laplacian of the xc-potential.}

In our algorithm for density inversion, the screening charge (the integral of the screening density over all space) is fixed; 
this stabilizes the minimization procedure and means 
we can constrain our potentials to be smooth and have the correct asymptotic behaviour, as we shall see that multiple potentials can arise from the inversion of the 
same density. 
Inverting DFT densities under appropriate constraints for the screening charge also provides a reliable procedure for alleviating self-interaction errors\cite{SIE_errors} 
in common DFT functionals.

The paper is structured as follows. In section \ref{Algorithm}, we demonstrate the algorithm used to minimize \eqref{eq:obju}. In section \ref{results}, we first demonstrate the 
accuracy and applicability of our method by inverting LDA densities for several molecules. We also show how inverting LDA densities under a constraint for the 
screening charge yields LDA potentials with self-interaction errors largely corrected. We then demonstrate how it can be applied to Hartree-Fock (HF) and 
coupled cluster densities to obtain accurate exchange-only and xc-potentials. Finally, we draw a brief comparison with the density inversion method of Zhao, 
Morrison and Parr\cite{Parr_inversion}, which uses the objective functional in Eq. \eqref{eq:obju} in a different manner.

\section{Method} \label{Algorithm}
In order to minimize the objective functional in \eqref{eq:obju}, we split the KS potential into the electron-nuclear part and an effective potential $v(\vec{r})$. 
{\color{black} At the minimum, the effective potential will coincide with} the Hxc-potential we seek, %$v_\textrm{Hxc}(\vec{r})$, 
for the KS system with density $\rho_t (\vec{r})$. 
We represent the effective potential $v(\vec{r})$ using a \emph{screening} density \cite{Gidopoulos_CLDA}:
\begin{align}
    v_s(\vec{r}) &= v_\textrm{en}(\vec{r}) + v(\vec{r}); \\
    v(\vec{r}) &= \int \dd{\vec{r}'} \frac{\rhoscr(\vec{r}')}{\rr}. \label{v_poisson}
\end{align}
This is always a valid representation for the potential due to Poisson's law \cite{Gorling_Poisson}. The screening density integrates to a screening charge $Q_\textrm{scr}$,
\begin{gather}
    \int \dd{\vec{r}} \rhoscr (\vec{r}) = Q_\textrm{scr}, \textrm{ with} \\
    N-1 \leq Q_\textrm{scr} \leq N.
\end{gather}
We argue that the value of $Q_\textrm{scr}$ is a measure of self-interactions (SIs) \cite{Gidopoulos_CLDA}: $Q_\textrm{scr}=N-1$ is a necessary condition 
for a method to be fully self-interaction free, otherwise the method is contaminated with self-interactions. As the value of $Q_\textrm{scr}$ does not change in the 
implementation of the method that we will describe, it is important to start with a screening density that is consistent with the screening charge of the target density.

When we vary $v(\vec{r})$ as $v(\vec{r})\to v(\vec{r})+\epsilon \, \delta v(\vec{r})$, with 
$\delta v (\vec{r}) = \int \dd{\vec{r}'} \delta \rho_{\rm scr} (\vec{r}')/|\vec{r}-\vec{r}'|$, the change in the Coulomb energy $U$ (functional of $v$) is given by
\begin{gather}
   \delta U[v] = \epsilon \iint\dd{\vec{r}} \dd{\vec{r}'} \delta\rhoscr(\vec{r})\tilde{\chi}_v(\vec{r},\vec{r}') \delta\rho(\vec{r}') + \mathcal{O}(\epsilon^2); \\
   \label{eq:rhodiff}
    \textrm{with } \delta \rho(\vec{r}) = \rho_v(\vec{r}) - \rho_t(\vec{r});  \\
    \textrm{and } \tilde{\chi}_v(\vec{r},\vec{r}')= \iint\dd{\vec{x}} \dd{\vec{y}} \frac{\chi_v(\vec{x}, \vec{y})}{|\vec{r}-\vec{x}| |\vec{r}'-\vec{y}|},
\end{gather}
where $\chi_v(\vec{r},\vec{r}')$ is the density-density response function for the KS system,
{\color{black}
\begin{equation}
    \chi_v(\vec{r},\vec{r}') = \sum_{i}^\textrm{occ}\sum_{a}^\textrm{unocc} \frac{\phi_{v,i}(\vec{r}) \phi_{v,a}^*(\vec{r}) \phi_{v,i}^*(\vec{r}')\phi_{v,a}(\vec{r}')}{\epsilon_{v,i}-\epsilon_{v,a}} + \textrm{c.c.}
\end{equation}
$\phi_{ v , i}$, $\phi_{ v , a}$ and $\epsilon_{ v , i}$, $\epsilon_{ v , a}$ are the occupied, unoccupied KS orbitals and their 
KS eigenvalues in the KS determinant with density $\rho_v$ (the ground state of $H_v$ in \eqref{ksv}). }

Since $\chi_v(\vec{r},\vec{r}')$ is a negative-semidefinite operator, if we vary $ \rhoscr(\vec{r})$ in the direction
\begin{equation}
    \rhoscr(\vec{r}) \to \rhoscr(\vec{r}) + \epsilon\delta\rho(\vec{r}),   \  \textrm{with} \ \epsilon>0, 
\end{equation}
then $U$ will decrease. 
We can therefore use a gradient-descent method to minimize $U$. 
This minimization will also ensure that the quantity $T_{\Psi}[v]$ in \eqref{eq:Tdiff} is minimized, since the functional derivative of $T_{\Psi}[v]$ \cite{Gidopoulos_1}
is equal to $- \delta \rho (\vec{r})$, when $\rho_t(\vec{r})$ is the density of $\Psi$. %=\rho_{\Psi}(\vec{r})$.

We note that during the minimization procedure, 
the screening charge $Q_\textrm{scr}$ remains equal to the value of the initial guess for $\rhoscr (\vec{r})$, since $\int \dd{\vec{r}} \delta\rho(\vec{r})=0$. 

\subsection{Algorithm} \label{sec:algorithm}

The method has been implemented in the Gaussian basis set code HIPPO  \footnote{For information, contact NL at lathiot@eie.gr. One- and two-electron integrals for the Cartesian Gaussian basis elements were calculated using the GAMESS code\cite{gamess1,gamess2}}.  The algorithm is described below.

\begin{enumerate}
    \item Initialize the screening density as follows:
    \begin{equation}
        \rhoscr^{(0)}(\vec{r}) = \frac{N-\alpha}{N} \rho^{(0)}(\vec{r}),
    \end{equation}
    where $\alpha \in [0,1]$ depends on the target density, and thus $Q_\textrm{scr}=N-\alpha$. $\rho^0(\vec{r})$ can be any density for the $N$-electron system. 
    
    $\rhoscr(\vec{r})$ is expanded in an auxiliary basis set,
    \begin{equation}
        \rhoscr(\vec{r}) = \sum_{k} \rho^{\textrm{s}}_{k} \theta_k(\vec{r}) .  \label{rho_s_orbs}
    \end{equation}
For our auxiliary basis we employed the density-fitted basis set\cite{density_fitting} corresponding to the orbital basis. 
Justification for this choice of auxiliary set is given in Appendix A. %the Supporting Information.
    \item Solve the single-particle KS equations,
    \begin{equation}
        \left [-\frac{\nabla^2}{2}+v_\textrm{en}(\vec{r})+v(\vec{r}) \right] \phi_{v , i} (\vec{r}) = \epsilon_{ v , i }  \, \phi_{v , i} (\vec{r}),
    \end{equation}
    to update the density $\rho_v(\vec{r})$.
    \item Update the screening density of the $i$-th iteration in the direction
    \begin{equation}
        \delta \rhoscr^{(i)} (\vec{r}) = \epsilon \, \big[ \rho_v^{(i)}(\vec{r}) - \rho_t(\vec{r}) \big],
    \end{equation}
    where $\epsilon$ is chosen with a quadratic line search to minimize $U$.
    
At this step, it is convenient for the target density to be expanded in the same basis set as the KS density $\rho_v(\vec{r})$, 
since the density difference is thus directly obtained. 
    \item Repeat steps 2 and 3 until either:
    \begin{enumerate}[i]
        \item $U$ and $\delta U$ are converged to within some chosen tolerances, or;
        \item The amount and rate of increase of \emph{negative} screening charge $Q_\textrm{neg} \ge 0$ exceeds a chosen amount, where
        \begin{equation} \label{nonp}
            Q_\textrm{neg} = \frac{1}{2} \left[
            \int\dd{\vec{r}} |\rho_\textrm{scr}(\vec{r})| - Q_\textrm{scr}  \right].
        \end{equation}
    \end{enumerate}
\end{enumerate}
Condition 4.ii is a kind of regularization \cite{Wasserman_review,Yang_regularization}. Due to both numerical issues (such as the effect of finite basis sets 
\cite{OEP_non_uniqueness,staroverov_2006_oep,Gidopoulos_OEP}), 
and possible theoretical constraints (non-interacting $v$-representability 
\cite{levy6062,kohn_v_representability,chen_stott_1991,chen_stott_1993,Godby_representability,dane_gonis}), 
converging $U$ to within the above tolerances can lead to spurious 
oscillations in the potential. This behaviour frequently coincides with a large build-up of negative screening charge, and thus a simple criterion to avoid these scenarios is to 
stop the procedure when this occurs. Details of the convergence criteria used can be found in Appendix B. %Supporting Information.

\section{Results} \label {results}
\subsection{Inversion of LDA densities} \label{sec:lda_rho_xc}
\begin{figure}
    \centering
    \includegraphics[width=\columnwidth]{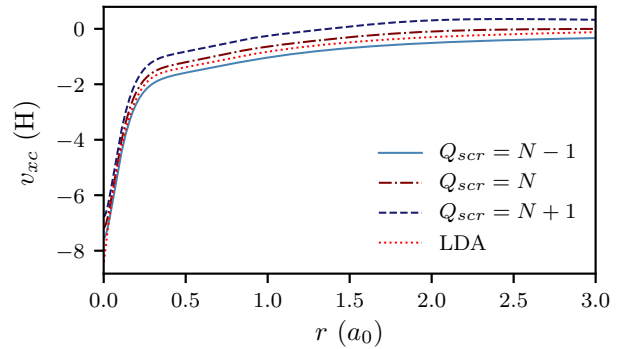}
    \caption{The inverted xc-potentials from the LDA density of Neon (cc-pVTZ), for different values of $Q_\textrm{scr}$. 
    Each value of $Q_\textrm{scr}$ produces a unique xc-potential.}
    \label{Ne_ldas}
\end{figure}

To demonstrate the applicability of our method, we first present results for the inversion of LDA densities for a few atomic and molecular systems. As previously discussed, it is important to begin with the correct $Q_\textrm{scr}$ for the system under consideration. As can be seen in Fig.~\ref{Ne_ldas}, minimizing 
$U [\rho_v - \rho_t] $ for the same target density yields a unique potential for every value of $Q_\textrm{scr}$.
Obviously, only the potential with the correct $Q_{\rm scr}$ will yield the target density $\rho_t$ exactly.

Since LDA potentials are contaminated with self-interactions, we would expect physically that $Q_\textrm{scr}=N$ in this case. However, this turns out not to be true 
when we transform from a grid-representation for the LDA xc-potential (as is typical in most codes), to the representation given by Eqs. \eqref{v_poisson} 
and \eqref{rho_s_orbs}. We observe that, in this representation, $Q_\textrm{scr}\neq N$  and is basis-set dependent. To determine the value of 
$Q_\textrm{scr}$, we solve the equation
\begin{gather}
    \rho^\textrm{xc}_{k} = \sum_{l} \ip*{\tilde{\theta}_k}{\theta_l}^{-1} \ip{\theta_{l}}{v_{xc}}, \textrm{with} \label{eq:rho_xc_exact}\\
    \rho_\textrm{xc}(\vec{r})  = \sum_k {\rho}^\textrm{xc}_{k} \theta_k(\vec{r}) , \ \ 
     \tilde{\theta}_k(\vec{r})  = \int \dd{\vec{r}'} \frac{\theta_k(\vec{r}')}{|\vec{r}-\vec{r}'|}.
\end{gather}
Here, $ \rho_{\rm xc}(\vec{r})$ is the effective xc-screening density, with $\int \dd{\vec{r}} \rho_{\rm xc}(\vec{r})=-\alpha$. Table \ref{He_Be_scr_chrgs} shows some values of $Q_\textrm{scr}$ for Helium and Beryllium with increasing basis set size. 

If desired, it is possible to approach $Q_\textrm{scr}=N$ by adding diffuse $s$-functions to the auxiliary basis set. 
As this only affects the potential by a small amount in the asymptotic tail, we choose not to modify the 
established basis sets in this work.

\begin{table}
\begin{ruledtabular}
    \centering
    \begin{tabular}{ccccc}
          & \multicolumn{2}{c}{He} &  \multicolumn{2}{c}{Be}\\
          \hline
          & $\alpha$ & IP (eV) & $\alpha$ & IP (eV) \\
          cc-pVDZ & 0.479 & 15.15 & 0.207 & 4.50\\
          cc-pVTZ & 0.214 & 14.82 & 0.148 & 4.81 \\
          cc-pVQZ & 0.301 & 15.41 & 0.185 & 5.29 \\
          cc-pV5Z & 0.256 & 15.89 & 0.165 & 5.41
    \end{tabular}
    \end{ruledtabular}
    \caption{Values of $\alpha$, where $Q_\textrm{scr}=N-\alpha$, and ionization potentials (IPs) as the negative of the HOMO energies, for He and Be with increasing basis set size. Basis sets are from Ref. \onlinecite{bse}.}
    \label{He_Be_scr_chrgs}
\end{table}

\begin{figure}
\centering
    \begin{subfigure}{\linewidth}
        \includegraphics[width=\columnwidth]{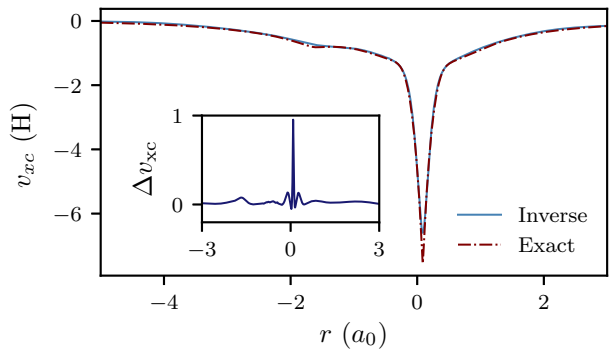}
        \caption{\ch{HF} (cc-pVTZ)}
    \end{subfigure}
    \begin{subfigure}{\linewidth}
        \includegraphics[width=\columnwidth]{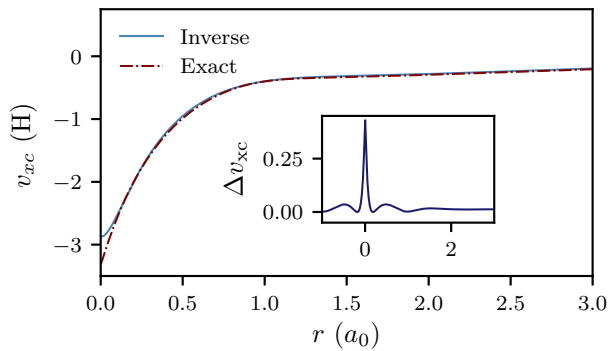}
        \caption{\ch{Be} (cc-pVQZ)}
    \end{subfigure}
    \caption{Comparison of xc-potentials for the inverted LDA density, and the exact LDA result.}
        \label{HF_Be_LDA_pots}
\end{figure}

With a method to calculate the appropriate value of $Q_\textrm{scr}$ for LDA densities, we now demonstrate the accuracy of our method when applied to LDA densities and the convergence with increasing basis set size. In Fig.~\ref{HF_Be_LDA_pots}, we see the qualitative similarities between the xc-potential from the inverted LDA density, and the actual LDA xc-potential. The region of biggest difference is observed near the nuclei; if accuracy in this region is desired, it is important to use a large basis set.

We can also use the HOMO energy as an indicator of the quality of the inversion procedure. In Table \ref{LDA_IP_errors}, we present results for the percentage difference between the actual and inverted HOMO energy for some atoms and molecules. These results demonstrate the improved accuracy with respect to basis set size, as well as a rough indication of how accurate we can expect our potentials to be with a given basis set. 

\begin{table*}
\begin{ruledtabular}
\begin{tabular}{cccccccccc}
    & \multicolumn{3}{c}{cc-pVDZ} & \multicolumn{3}{c}{cc-pVTZ} & \multicolumn{3}{c}{cc-pVQZ} \\
    \hline
    IP (eV) & Inverse & LDA & \% err & Inverse & LDA & \% err & Inverse & LDA & \% err \\
    \hline
    \ch{He} & 15.15 &15.14 & 0.1 & 14.82 & 15.47 &4.2  &15.41	&15.37 &	0.6 \\
\ch{Be} &	4.50	&5.62&		19.9&	4.81&	5.60&	14.1&	5.29	&5.60&	5.5 \\
\ch{Ne} &	6.69	&12.24	&	45.3&	10.56	&13.17&	19.8&	11.75&	13.40&	12.3 \\
\ch{HF}	& 7.18	& 8.45	&15.0	&8.91&	9.38&	5.0&	9.37&	9.64&		2.8\\
\ch{H2O}& 5.71&	6.23	&	8.3 &    6.67	&7.00	&4.7&	6.86	&7.21	&4.4 \\
\ch{H2} &	9.53&	10.12&		5.8& 	10.00&	10.25	&2.4&	10.02&	10.26	&2.3\\
\ch{CO} &	6.16&	8.71&		29.3&	7.73&	9.07&	14.8&	8.82&	9.11&	3.2 \\
Avg \% err	& - &- &		17.7 & - & - &	9.3 & - & -	&	4.5

\end{tabular}
\end{ruledtabular}
    \caption{Comparison of IPs (from HOMO energies) of the inverted LDA densities with the actual LDA IPs.}
    \label{LDA_IP_errors}
\end{table*}

\subsection{Constrained LDA results}

In the previous subsection, we demonstrated the importance of choosing the right screening charge when inverting LDA densities. However, the flexibility we have in choosing the screening charge can be used to our advantage, to remove the effects of self-interactions (SIs) from LDA and other SI contaminated densities by setting $Q_\textrm{scr}=N-1$. The success of this `constrained DFT' approach has been already demonstrated \cite{Gidopoulos_CLDA,Pitts2018}, but using a different method in which the energy is minimized under the following constraints:
\begin{align}
    Q_\textrm{scr}&=N-1, \ \textrm{ and} \\
    \rho_\textrm{scr} (\vec{r})&\ge 0. \label{strongp}
\end{align}
The second constraint \eqref{strongp} is an approximation, which in the aforementioned method is required to prevent a negative screening charge `hole' localizing at infinity. 
%We have not employed this positivity constraint in our density inversion approach. 
In our density inversion approach we have employed the weaker condition 4.ii (\ref{nonp}) instead of \eqref{strongp}.
%This is useful as, by comparing results between the two methods, we can test the validity of the positivity constraint and have the option of 
%an alternative approach in cases where it is inaccurate.

In Table \ref{CLDA_IPs}, we see a comparison of the ionization potentials (IPs), taken to be the negative of the HOMO orbital energies \cite{IP_thm}. 
We see that inverting the density under the constraint $Q_\textrm{scr}=N-1$, and our previous constrained-LDA (CLDA) method \cite{Gidopoulos_CLDA} 
with the positivity constraint, both yield very similar results for the IPs. 
As discussed in earlier work and seen here, this constrained method yields consistently better IPs than normal LDA, but preserves the energetics from the 
LDA calculation. 
Further analysis of the tendency for $\rho_\textrm{scr}(\vec{r})$ to be positive can be found in Appendix A. %the supporting information.

\begin{table}
\begin{ruledtabular}
    \begin{tabular}{ccccc}
         &  LDA & CLDA (inv) & CLDA \cite{Gidopoulos_CLDA} & Expt. \cite{Nist_data} \\
         \hline
         \ch{He} & 15.47 & 23.12 & 23.82 & 24.59 \\
         \ch{Be} & 5.60 & 8.48 & 8.65 & 9.32 \\
         \ch{Ne} & 13.17 & 18.85 & 18.89 & 21.56\\
         \ch{HF} & 9.38 & 14.08 & 14.17 & 16.03\\
         \ch{H2O} & 6.83 & 11.10 & 11.04 & 12.62\\
         \ch{H2} & 10.25 & 15.15 & 15.64 & 15.43 \\
         \ch{CO} & 8.97 & 12.50 & 12.84 & 14.01\\
    \end{tabular}
    \caption{Comparison of IPs (from HOMO energies) for constrained-LDA using the inversion of density, and our previous CLDA method \cite{Gidopoulos_CLDA}. 
    All basis sets are cc-pVTZ.}
    \label{CLDA_IPs}
    \end{ruledtabular}      
\end{table}

\subsection{Inversion of `non-local' densities}

The principal application of the density inversion scheme is to invert densities obtained with non-DFT methods to find the KS potential which shares the same density. 
We have applied our scheme to two densities calculated with Hartree-Fock (HF) and Coupled Cluster (CCSD(T)) theories, with target CCSD(T) densities obtained from the PSI4 code \cite{Psi4_main,Psi4_CC}. We focus on these because the inversion of an HF density gives us an exchange-only local potential in DFT 
(local Fock exchange, LFX \cite{Hollins_LFX}), which is a close approximation to the exact-exchange potential \cite{Staroverov_ELPs,Hollins_LFX}.
CCSD(T) calculations yield highly accurate densities \cite{CC_intro}, which give us an idea of what the `exact' xc-potential in KS theory should be.

Just as for the LDA case, it is important to choose the correct value for the screening charge. As both HF and CCSD(T) are self-interaction free, we expect 
$Q_\textrm{scr}=N-1$. Unlike in the LDA case, there is no way of determining if this is the exact numeric value; however, our results strongly suggest this is 
a good choice. We again focus on the IPs obtained from the HOMO orbital energies to judge the quality of our inversion procedure. For HF-inverted densities, by 
Koopmans' theorem \cite{KOOPMANS1934104} and its analogue in DFT relating the HOMO energy to the IP\cite{IP_thm}, we expect the inverted $\epsilon_\textrm{H}$ 
to equal $\epsilon_\textrm{H}$ from HF. Meanwhile, for the densities inverted from CCSD(T), the difference in the IP compared to experiment should offer insight into the 
reliability of the procedure.

In table \ref{HF_IPs}, we see how the IPs taken from the HOMO energies of the inverted local potential compare with the IPs from HF theory. These results indicate what level of accuracy can be expected with a given basis set: it appears we should use at least cc-pVTZ basis sets to obtain an accurate potential, with an average difference of 3.4\% between the inverted and actual IPs. More accurate results can be obtained if desired by increasing the basis set size. A similar picture emerges for the inverted CCSD(T) densities, as seen in Table \ref{CC_IPs}; in this case, cc-pVQZ results are not computed due the expense of obtaining the coupled cluster density matrix for these densities, but we see a very similar result for the average error in cc-pVTZ basis sets.

\begin{table*}
\begin{ruledtabular}
\begin{tabular}{cccccccccc}
    & \multicolumn{3}{c}{cc-pVDZ} & \multicolumn{3}{c}{cc-pVTZ} & \multicolumn{3}{c}{cc-pVQZ} \\
    \hline
    IP (eV) & Inverse & HF & \% err & Inverse & HF & \% err & Inverse & HF & \% err \\
    \hline
    \ch{He} &25.23&	24.88&	1.4	&24.97&	24.97&	0.0 &24.98 &	24.98	&0.0 \\
\ch{Be}& 8.96&	8.41&	6.5&	8.42&	8.42&	0.0&	8.37&	8.42&	0.6 \\
\ch{Ne}& 17.57&	22.65&	22.4	&22.19&	23.01&	3.6	&24.40&	23.10&	5.6 \\
\ch{HF} &14.21	&17.12	&17.0	&16.57	&17.52&	5.4	&17.23&	17.64	&2.3\\
\ch{H2O} &12.03&	13.44&	10.5&	12.99&	13.76&	5.6&	13.40&	13.85&	3.2 \\
\ch{H2} & 16.13	&16.10	&0.2&	16.16&	16.16	&0.0	&16.17&	16.17	&0.0 \\
\ch{CO} & 11.65&	14.96&	22.1	&13.74	&15.09	&8.9	&14.03&	15.11	&7.1\\
Avg \% err	& - &- &		11.5 & - & - &	3.4 & - & -	&	2.7

\end{tabular}
\end{ruledtabular}
    \caption{Comparison of IPs for the local potential of an HF density with the actual HF IPs.}
    \label{HF_IPs}
\end{table*}

\begin{table}
\begin{ruledtabular}
    \begin{tabular}{cccccc}
         &  \multicolumn{2}{c}{cc-pVDZ} & \multicolumn{2}{c}{cc-pVTZ}\\
         \hline
         IP (ev) & Inverse & \% err & Inverse & \% err & Expt \cite{Nist_data}\\
         \hline
    \ch{He} & 24.94	 & 1.4 &24.57 &	0.1 & 24.59 \\
\ch{Be}	& 9.13 & 2.0 & 9.12 & 2.0 &	9.32 \\
\ch{Ne} & 12.09 & 43.9 & 20.41	& 5.3 &	21.56 \\
\ch{HF} & 11.34 & 29.3 & 15.43 & 3.7 &	16.03 \\
\ch{H2O} & 10.01 & 20.7 & 12.28 & 2.7 &	12.62 \\
\ch{H2} & 15.91	& 3.1 &	16.45 & 6.6 & 15.43 \\
\ch{CO}	 & 10.01 & 28.6 & 13.18 & 5.9 &	14.01 \\
Avg \% err & - & 18.4 &	- &	3.8 & - \\	
    \end{tabular}
    \end{ruledtabular}      
    \caption{Comparison of IPs for the local potential of a CCSD(T) density with experimental IPs.}
    \label{CC_IPs}
\end{table}

\begin{figure}
\centering
    \begin{subfigure}{\linewidth}
        \includegraphics[width=\columnwidth]{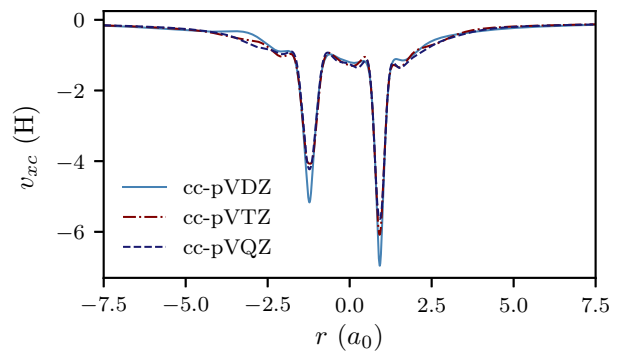}
        \caption{\ch{CO} (HF)}
    \end{subfigure}
    \begin{subfigure}{\linewidth}
        \includegraphics[width=\columnwidth]{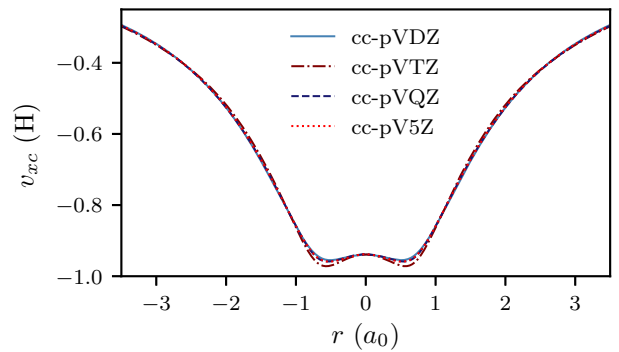}
        \caption{\ch{H2} (CCSD(T))}
    \end{subfigure}
    \caption{xc-potentials for (a) inverted HF density of \ch{CO}, (b) inverted CCSD(T) density of \ch{H2} for various basis sets.}
        \label{HF_CC_pots}
\end{figure}

Besides these IP comparisons, we demonstrate the applicability of our method by plotting some xc-potentials. In Fig. \ref{HF_CC_pots}, we see that the xc-potentials converge with basis set and produce smooth potentials. As in the LDA case, the inversion procedure struggles most in the regions very close to the nuclei. However, the inverted potentials appear to converge well for the purposes of qualitative analysis outside of these regions.

\begin{figure}
\includegraphics[width=\columnwidth]{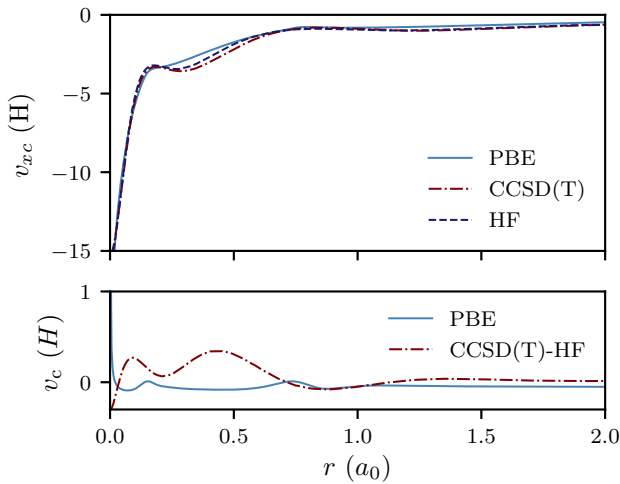}%
\caption{Top: \ch{Ar} (cc-pvTZ) xc-potentials from inverted HF and CCSD(T) densities, and PBE; bottom: correlation potentials, from the difference of CCSD(T) and HF inverted xc-potentials, and PBE.\label{Ar_HF_CC_PBE}}%
\end{figure}

We can also obtain approximate correlation potentials by taking the difference between the (almost) fully correlated inverted CCSD(T) potential, and the exchange-only inverted HF potential. We can expect this to yield accurate correlation potentials when the system under consideration is weakly-correlated, as in this case the inverted HF potential is close to the exact-exchange potential \cite{Hollins_LFX,Staroverov_ELPs}. In Fig. \ref{Ar_HF_CC_PBE}, we have plotted this correlation potential and the xc-potential for Argon, along with a comparison with the PBE potential.

\section{Comparison with the method by Zhao, Morrison, Parr}
Zhao, Morrison and Parr (ZMP), in their well-known density-inversion method \cite{Parr_inversion}, impose the constraint that the Coulomb energy 
\mbox{$U[\rho-\rho_t]$} \eqref{eq:obju} actually vanishes, rather than be minimised. 
The KS potential in their method, 
\begin{equation} \label{zmp}
v_s^\Lambda(\vec{r}) = v_\textrm{en} (\vec{r}) + \bigg(1 - {1 \over N}\bigg) v_\textrm{H} [ \rho ] (\vec{r})  
+ \Lambda \! \int \! \dd \vec{r}' \, {\rho (\vec{r}') - \rho_t (\vec{r}') \over \rr} ,
\end{equation} 
consists of the external potential $v_\textrm{en} (\vec{r})$, the Fermi-Amaldi potential $(1-1/N) \, v_\textrm{H} [\rho] (\vec{r})$, with 
$v_\textrm{H} [\rho] (\vec{r})$ the Hartree potential, and finally an effective potential to satisfy the constraint of zero \mbox{$U[\rho - \rho_t]$}, in the limit of diverging 
Lagrange multiplier $\Lambda \rightarrow \infty$.
ZMP argue that inclusion of the Fermi-Amaldi potential in their KS potential is auxiliary, to aid convergence and relieve the burden of the xc-potential when $\Lambda$ is finite. %; they claim that its omission still yields exact equations.
However, %our analysis reveals that, 
at any finite $\Lambda$, inclusion of the Fermi-Amaldi potential in \eqref{zmp} is crucial 
%to allow the effective Hxc potential in the ZMP method to provide the screening required to reproduce the target density.
since it is the term that provides the correct screening charge required by the target density. 
Its omission would imply that in the asymptotic region, a KS electron would be attracted by the full, unscreened nuclear charge.
{\color{black}See also the discussion by Liu, Ayers and Parr in Ref.~\onlinecite{Liu_parr}}.

The connection and similarity between the method by ZMP and ours is analogous to the connection between the direct minimisation 
of a total energy density-functional and its indirect minimisation using the optimised effective potential (OEP) method \cite{sharp_horton, talman_shadwick}.
The ZMP KS equations can be derived by the direct minimisation of the standard DFT total energy expression (as a density functional), 
using $E_\textrm{xc}^\textrm{ZMP} [ \rho ] = \Lambda U[ \rho - \rho_t] - (1/N) U [ \rho ]$ in place of the `xc' energy density-functional. 
%is given by $E_{xc}^{ZMP} [ \rho ] = \Lambda U[ \rho - \rho_t] - (1/N) U [ \rho ]$.
The total energy minimization must then be carried out for various values of $\Lambda$ and the results extrapolated to $\Lambda \rightarrow \infty$.  
The analogy with our method is that we only work with $\Lambda = \infty$ and rather than the whole total energy, we minimise just $U [ \rho - \rho_t]$. 
Only now, $U [\rho - \rho_t]$ becomes a functional of the effective potential $v_\textrm{en} + v$ that yields $\rho$, i.e., $\rho= \rho_v$, and $U [\rho_v - \rho_t]$ 
must be minimised with the OEP method.

\section{Discussion} \label{discussion}

We have presented a reliable inversion method to find the local KS potential corresponding to given target density. This method utilizes the concept of a screening density, which offers both a way of controlling the minimization procedure to yield physical potentials and also aids our understanding of self-interactions in DFT. 

The steepest descent method presented here is a stable method to invert the density and works well for large enough basis sets for atoms and molecules 
at their equilibrium geometries. 
Work is in progress to improve convergence for more complicated input densities (such as for stretched molecules) 
and will be presented in a future publication.

\begin{acknowledgements}
NIG acknowledges financial support by The Leverhulme Trust, through a Research Project Grant with number RPG-2016-005.\\
NNL acknowledges support by the project ``Advanced Materials and Devices'' (MIS 5002409) funded by
NSRF 2014-2020. \\
TJC and NIG thank Prof. Rod Bartlett for very helpful discussions during his visit at Durham University in early 2019 and acknowledge 
the Institute of Advanced Study at Durham University for hosting this visit. 
\end{acknowledgements}

\section{Data availability}

The data that support the findings of this study are available from the corresponding author upon reasonable request.

\appendix
\section{Choice of basis set representation for $\rho_\textrm{scr}$}

As discussed in \S \ref{sec:algorithm}, we expand the screening density in an auxiliary basis set which is the density-fitted set corresponding to the orbital basis. This is an intuitive choice, because we represent an effective density with a basis set designed for densities; it is also a convenient choice, because density-fitted sets are frequently used anyway to accelerate the computation of integrals in quantum Chemistry codes\cite{baerends1973self,EICHKORN1995652}.

To justify this choice quantitatively, we recall that we can obtain directly the Gaussian representation of the LDA grid potential using Eq. \eqref{eq:rho_xc_exact}. 
As a measure to gauge the quality of defining the potential in a given basis set, we use the Coulomb energy $U[\rho_\textrm{ga} - \rho_\textrm{gr}]$ \eqref{eq:obju},
where $\rho_\textrm{ga}$ and $\rho_\textrm{gr}$ are the densities arising from defining the potential in a Gaussian basis set and on the grid respectively. 
The smaller the value of $U[\rho_\textrm{ga} - \rho_\textrm{gr}]$, the better one might expect the Gaussian representation to be. In Table \ref{tab:U_values}, we compare
 values of $U[\rho_\textrm{ga} - \rho_\textrm{gr}]$ for three choices of basis function for the screening density: the orbital basis, the density-fitted basis, and also the 
 uncontracted orbital basis, which is a common choice for the potential\cite{Pitts2018,Gorling_uncontracted}. We observe that the density-fitted sets give the closest fit to the 
 grid representation based on this criterion.

\begin{figure}
\centering
    \begin{subfigure}{\linewidth}
        \includegraphics[width=\columnwidth]{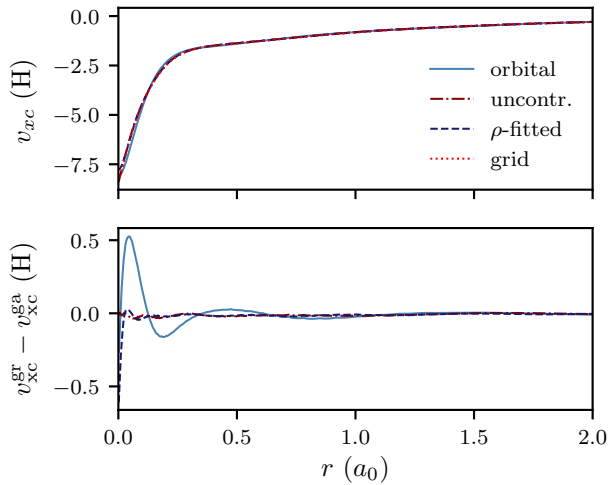}
        \caption{\ch{Ne} (cc-pVTZ)}
    \end{subfigure}
    \begin{subfigure}{\linewidth}
        \includegraphics[width=\columnwidth]{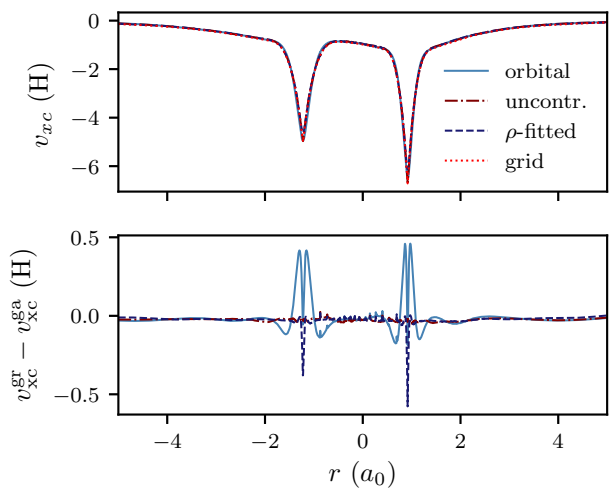}
        \caption{\ch{CO} (cc-pVTZ)}
    \end{subfigure}
    \caption{Comparison of the LDA xc-potential on a grid, against various Gaussian basis set representations. Lower images show the differences between the grid and Gaussian representation.}
        \label{fig:gaussian_comparisons}
\end{figure}

In Fig. \ref{fig:gaussian_comparisons}, we plot the LDA xc-potentials for these basis set choices. In contrast to the analysis above, the uncontracted sets seem to give the 
best fit to the grid potential, but we note that the density-fitted sets give a close fit everywhere except the nuclear positions. In our experience, the algorithm works more 
smoothly for the density-fitted sets than the uncontracted ones. Given that we minimize $U[\rho_v - \rho_t]$, it makes sense to choose a representation which also minimizes this 
expression. The gradient-descent algorithm also struggles to reproduce the target density near the nuclei regardless of the auxiliary basis chosen, so the lack of accuracy of 
the density-fitted sets in this region is not so important in our method.

\begin{table}[]
    \begin{ruledtabular}
    \begin{tabular}{cccc}
          $U[\rho_\textrm{ga} - \rho_\textrm{gr}]$ & orbital & uncontracted & $\rho$-fitted  \\
          \hline
          He & \num{2.3e-7} & \num{2.1e-7} & \num{1.2e-8} \bigstrut[t] \\ % He & \num{4.6e-7} & \num{4.1e-7} & \num{2.3e-8} \bigstrut[t] \\
          Be & \num{7.0e-4} & \num{5.5e-9} & \num{4.2e-10}\\ % Be & \num{1.4e-3} & \num{1.1e-8} & \num{8.3e-10}\\
          Ne & \num{9.0e-5} & \num{1.8e-6} & \num{3.4e-10} \\ % Ne & \num{1.8e-4} & \num{3.6e-6} & \num{6.8e-10} \\
          HF & \num{9.0e-5} & \num{2.9e-7} & \num{7.5e-9} \\ % HF & \num{1.8e-4} & \num{5.8e-7} & \num{1.5e-8} \\
          \ch{H2O} & \num{1.2e-4} & \num{2.2e-7} & \num{8.5e-9}\\ %  \ch{H2O} & \num{2.3e-4} & \num{4.4e-7} & \num{1.7e-8}\\
          \ch{H2} & \num{7.0e-8} & \num{1.6e-7} & \num{6.0e-8} \\ % \ch{H2} & \num{1.4e-7} & \num{3.1e-7} & \num{1.2e-7} \\
          \ch{CO} & \num{3.5e-4} & \num{2.7e-7} & \num{1.6e-9} % \ch{CO} & \num{7.0e-4} & \num{5.3e-7} & \num{3.1e-9}
    \end{tabular}
    \end{ruledtabular}
    \caption{Values of $U[\rho_\textrm{ga} - \rho_\textrm{gr}]$ for LDA potentials in different Gaussian basis sets. All bases cc-pVTZ.}
    \label{tab:U_values}
\end{table}

\section{Convergence criteria}

The convergence criteria for the objective functional $U$ and the change in objective functional $\delta U$ were set to $5\times10^{-9}$ Hartree and $5\times10^{-11}$ Hartree per electron respectively. If both of these conditions are satisfied, $U$ is taken to be converged.

In general, satisfying the above criteria is not a problem when inverting a DFT density (eg LDA). However, when inverting non-local densities, the problem of spurious oscillations tends to emerge and thus it is necessary to use a regularization criterion. As mentioned in \S \ref{sec:algorithm}, we monitor the amount of negative screening charge to indicate the onset of these spurious oscillations.

The onset of negative screening charge is dependent on several factors, including:
\begin{enumerate}[i]
    \item the number of electrons $N$;
    \item the size of the basis set;
    \item the target density;
\end{enumerate}
and other (hard to quantify) factors relating to the system under consideration. To guide our intuition, we use the procedure outlined in \S \ref{sec:lda_rho_xc} to determine the behaviour of the `exact' $\rho_s(\vec{r})$ for LDA densities.

\begin{figure}
\centering
    \begin{subfigure}{\linewidth}
        \includegraphics[width=\columnwidth]{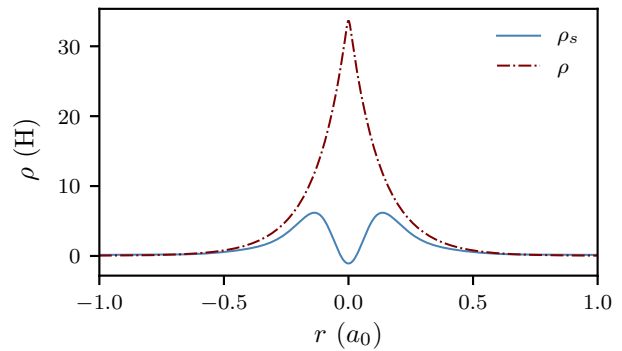}
        \caption{\ch{Be} (cc-pVTZ)}
    \end{subfigure}
    \begin{subfigure}{\linewidth}
        \includegraphics[width=\columnwidth]{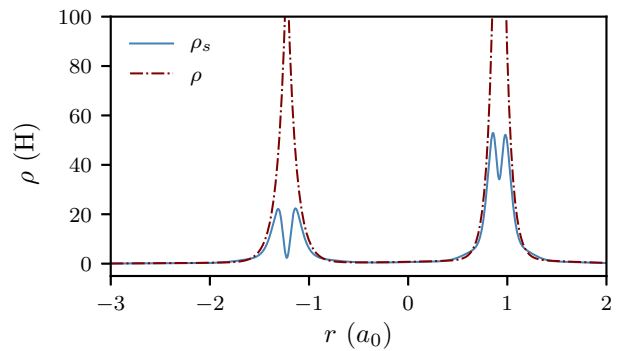}
        \caption{\ch{CO} (cc-pVTZ)}
    \end{subfigure}
    \caption{Effective screening densities, $\rho_s(\vec{r})$, for LDA densities, with the actual densities for comparison. We obesrve the tendency for a small amount of negative screening charge near the nuclei.}
        \label{LDA_rho_s_plots}
\end{figure}

In Table \ref{tab:rho_neg_lda}, we see that a small amount of negative screening charge is typically present for the LDA effective screening density. In Fig. \ref{LDA_rho_s_plots}, we see this negative screening density has a tendency to build up near the nuclei. There is no reason to expect dramatically dissimilar behaviour for different target densities, and therefore it seems judicious to allow a small amount of negative screening charge to manifest itself in the inversion procedure. However, as previously discussed, if $Q_\textrm{neg}$ is permitted to increase too fast or become too large, then we observe the onset of undesirable oscillations in the potential.

\begin{table}[]
    \begin{ruledtabular}
    \begin{tabular}{ccc}
          $Q_\text{neg}$ & cc-pVDZ & cc-pVTZ \\
          \hline
          He & 0.0 & \num{9.88e-3} \\
          Be & \num{5.81e-2} & \num{7.65e-2} \\
          Ne & 0.0 & \num{3.30e-4} \\
          HF & \num{4.50e-2} & \num{8.18e-2} \\
          \ch{H2O} & \num{3.03e-2} & \num{1.15e-1} \\
          \ch{H2} & \num{6.55e-3} & \num{6.35e-2} \\
          \ch{CO} & \num{1.09e-2} & \num{3.51e-4}
    \end{tabular}
    \end{ruledtabular}
    \caption{Amount of negative screening charge, $Q_\textrm{neg}$, for exact LDA screening densities.}
    \label{tab:rho_neg_lda}
\end{table}

With the above arguments in mind, we monitor the following variables during the inversion procedure:
\begin{enumerate}[i]
    \item Soft limit, $Q_\textrm{neg}^\textrm{soft}$;
    \item Change in $Q_\textrm{neg}$, $\delta Q_\textrm{neg}$ between iterations;
    \item Hard limit, $Q_\textrm{neg}^\textrm{hard}$;
\end{enumerate}
If both conditions (i) and (ii) are satisfied, or just condition (iii), the calculation stops. For all the results published in this paper, we use the same values which are equal to:
\begin{enumerate}[i]
    \item $Q_\textrm{neg}^\textrm{soft}=0.01$;
    \item $\delta Q_\textrm{neg}=0.005$;
    \item $Q_\textrm{neg}^\textrm{hard}=0.05$;
\end{enumerate}
where all the above values are quoted per electron. The above values give reasonable results for the systems presented in this paper, which are all atoms or molecules at 
their equilibrium geometries. However, we have observed that for molecules stretched beyond their equilibrium geometries, a large build-up of negative screening charge 
develops. A more sophisticated procedure would be required for these and other difficult cases.

%\bibliography{references.bib}


\end{document}